\documentclass{SciPost}

\hypersetup{
    colorlinks,
    linkcolor={red!50!black},
    citecolor={blue!50!black},
    urlcolor={blue!80!black}
}

\usepackage[bitstream-charter]{mathdesign}
\usepackage{bm}
\usepackage{physics}

\DeclareSymbolFont{usualmathcal}{OMS}{cmsy}{m}{n}
\DeclareSymbolFontAlphabet{\mathcal}{usualmathcal}

\fancypagestyle{SPstyle}{
\fancyhf{}
\lhead{\colorbox{scipostblue}{\bf \color{white} ~SciPost Physics }}
\rhead{{\bf \color{scipostdeepblue} ~Submission }}

\fancyfoot[C]{\textbf{\thepage}}
}

\newcommand{\eref}[1]{Eq.~\eqref{#1}}
\newcommand{\srf}[1]{Sec.~\ref{#1}}
\newcommand{\arf}[1]{App.~\ref{#1}}
\newcommand{\frf}[1]{Fig.~\ref{#1}}
\newcommand{\ee}{\mathrm{e}}
\newcommand{\Pimp}{\hat{P}_{\rm imp}}
\newcommand{\kF}{k_{\rm F}}
\newcommand{\vF}{v_{\rm F}}
\newcommand{\EF}{\varepsilon_{\rm F}}
\newcommand{\Lam}{\Lambda}
\newcommand{\Lamt}{\widetilde{\Lambda}}
\newcommand{\Et}{\widetilde{E}}
\newcommand{\sgn}{\operatorname{sgn}}
\newcommand{\Erf}{\operatorname{erf}}
\newcommand{\FS}{|{\rm FS}\rangle}
\graphicspath{{figs/}{./}}

\begin{document}

\pagestyle{SPstyle}

\begin{center}{\Large \textbf{\color{scipostdeepblue}{
Relaxation dynamics of the impurity momentum in a one-dimensional Fermi gas
}}}\end{center}

\begin{center}\textbf{
Felipe Taha Sant'Ana\textsuperscript{1$\star$}
}\end{center}

\begin{center}
{\bf 1} Center of Mathematics, Computing, and Cognition, Federal University of ABC,
09210-580 Santo Andr\'e, Brazil
\\[\baselineskip]
$\star$ \href{mailto:t.felipe@ufabc.edu.br}{\small t.felipe@ufabc.edu.br}
\end{center}

\section*{\color{scipostdeepblue}{Abstract}}
{\boldmath\textbf{%
We consider a single mobile impurity coupled by a
repulsive contact interaction to a gas of free spinless fermions of
equal mass in one dimension. Using the determinant overlaps between the
interacting eigenstates and free-fermion states, we derive closed
determinant formulas for the matrix elements of the impurity momentum
operator and of its square between arbitrary Bethe-ansatz eigenstates
of a finite system. At infinite coupling we prove a selection rule:
both operators have vanishing matrix elements between distinct
eigenstates that share the same rapidity, and every eigenstate carries
the per-particle share of the total momentum. For an impurity injected
into the Fermi sea we have the following: at infinite coupling the
average impurity momentum relaxes completely in the thermodynamic
limit, while at finite coupling relaxation remains incomplete. The
closed-form matrix elements turn the spectral representation of the
average impurity momentum into explicit finite sums over eigenstates.
We evaluate the post-injection dynamics, resolve the quantum-flutter
oscillations and the approach towards the diagonal-ensemble value, and
map the stationary momentum as a function of the injection momentum
and the coupling. Finally, we analyse the thermodynamic limit
and its asymptotics as an integral over Fredholm determinants.
}}

\vspace{10pt}
\noindent\rule{\textwidth}{1pt}
\tableofcontents
\noindent\rule{\textwidth}{1pt}
\vspace{10pt}

\section{Introduction}
\label{sec.intro}

In a mobile impurity moving through a one-dimensional (1D) quantum
liquid, many-body relaxation can be followed in real time. Cold-atom
experiments can inject a distinguishable
atom into a 1D gas with a well-defined velocity and follow its momentum
distribution
\cite{Palzer2009,Catani2012,Wenz2013,Meinert2017,Dhar2025,Horvath2026}.
On the theory side, kinematic constraints of 1D motion invalidate the
quasiparticle picture familiar from three-dimensional polarons
\cite{Massignan2014,GrusdtDemler2015,Grusdt2025} and produce phenomena
with no higher-dimensional counterpart \cite{Cazalilla2011,Imambekov2012}.
The dispersion of the dressed impurity becomes periodic in its momentum,
with a period set by the density of the host \cite{Lamacraft2009,Schecter2012};
a driven impurity therefore performs Bloch oscillations without a
lattice \cite{GangardtKamenev2009,Schecter2012,Meinert2017,MajumdarPetkovic2025,Scopa2026};
and at small momentum the impurity motion is governed by the infrared
excitations of the host, beyond any quasiparticle description
\cite{Zvonarev2007,Kantian2014}. Fast impurities show their own dynamical
transitions, such as the quantum Cherenkov transition
\cite{Seetharam2021,Seetharam2024}.
Two of these phenomena motivate the present work. First,
quantum flutter: an impurity injected with momentum comparable to
or exceeding the Fermi momentum sheds momentum through
long-lived oscillations \cite{Mathy2012,Knap2014,Dolgirev2021}. Second,
incomplete relaxation: in an integrable system the impurity
momentum saturates at a nonzero stationary value, so that a portion of
the initial momentum survives forever
\cite{Mathy2012,Lychkovskiy2015}, a behaviour
recently observed directly \cite{Horvath2026}.

The minimal integrable model of this physics was solved by McGuire
\cite{McGuire1965,McGuire1966}: a single impurity of mass equal to the
host fermions, coupled to them by a repulsive contact interaction. It is
the one-magnon sector of the Yang--Gaudin gas
\cite{Yang1967,Gaudin1967,Guan2013}, and most analytic results on 1D
impurity dynamics build on it: spectral properties
\cite{CastellaZotos1993,McGuire1966}, the impurity Green's function as a
Fredholm determinant \cite{GPZ2015}, and its finite-temperature
dynamics \cite{GPS2022,GPS2023}. Lattice versions of the problem admit
exact calculations of the polaron spectral function
\cite{HuWangLiu2025,LiuHu2025}. The studies that identified quantum
flutter evaluated the impurity momentum by summing matrix elements
obtained numerically from Bethe-ansatz form factors
\cite{Mathy2012,Knap2014}, and a Bethe-ansatz analysis of the
supersonic regime in a bosonic host was given in
Ref.~\cite{Zhang2023}.

Motivated by the fact that no closed-form expression for the matrix
elements of the impurity momentum between McGuire eigenstates has been
available, nor a determinant representation analogous to the one known
for the Green's function, we develop a description of the
time-dependent average momentum. Our starting point is the determinant
formula for the overlaps between interacting eigenstates and free
states \cite{GPZ2015}. Then, we obtain the matrix elements of the impurity momentum and
of its square between arbitrary eigenstates of a finite system, each as
a single determinant times elementary factors. The diagonal
elements reproduce the known polaron results.

At infinite coupling the spectrum organizes into sectors of eigenstates
that share the same rapidity. We prove that the impurity momentum and
its square have no matrix elements between distinct states of the same
sector, and that every eigenstate carries the per-particle share of the
total momentum. After injection the mean impurity momentum therefore
relaxes completely in the thermodynamic limit, and the
incomplete-relaxation plateau \cite{Mathy2012,Lychkovskiy2015}
disappears at infinite coupling.

The closed matrix elements turn the spectral sums for the average
momentum and its variance into finite sums of products of
determinants. With them we follow the post-injection dynamics in finite
systems, resolve the flutter oscillations at strong coupling and the
approach towards the diagonal-ensemble value, and map the stationary
momentum as a function of the injection momentum and the coupling. In
the thermodynamic limit we derive a representation of the average
momentum as a double rapidity integral over Fredholm determinants of
explicit kernels acting on the Fermi sea. At infinite coupling the
kernels are built from error functions, and we show analytically how the
initial momentum emerges from this representation and determine its
long-time structure.

The paper is organized as follows. \srf{sec.model} recalls the model,
its Bethe-ansatz solution and the overlap formula, and fixes the
injection protocol. \srf{sec.matrix-elements} presents the closed
matrix elements, the special limits, and the infinite-coupling
selection rules. \srf{sec.dynamics} sets up the spectral representation
of the average momentum, establishes the diagonal-ensemble limit and the
infinite-coupling statements, and derives the Fredholm-determinant
representation. \srf{sec.numerics} presents the relaxation dynamics
in finite systems.
\srf{sec.conclusions} concludes. Technical material is collected in the
appendices: the determinant reduction of the generating function
(\arf{app:cauchy-binet}), its small-$y$ expansion, which yields the
matrix elements (\arf{app:smally}), the degenerate-sector analysis
(\arf{app:degenerate}), the derivation of the determinant
representation of the average momentum (\arf{app:pt}), and its
infinite-coupling kernels (\arf{app:alpha0}).

\section{Model, spectrum, and overlaps}
\label{sec.model}

We consider $N$ free spinless fermions of unit mass and one
distinguishable particle (``the impurity'') of the same mass on a ring
of circumference $L$ ($\hbar=1$),
\begin{equation}
\label{eq.H}
H = -\frac12\frac{\partial^2}{\partial X^2}
-\frac12\sum_{j=1}^{N}\frac{\partial^2}{\partial x_j^2}
+ g\sum_{j=1}^{N}\delta(X-x_j) ,
\qquad g>0 ,
\end{equation}
where $X$ is the impurity coordinate and $x_j$ are those of the gas
particles. The gas is characterized by the density
$\rho=N/L$, the Fermi momentum $\kF=\pi\rho$, the Fermi energy
$\EF=\kF^2/2$ and the Fermi velocity $\vF=\kF$. The dimensionless
coupling is
\begin{equation}
\gamma=\frac{g}{\rho},
\qquad
\alpha\equiv\frac{2}{g}=\frac{2}{\gamma\rho} .
\end{equation}
The model is integrable \cite{McGuire1965,Yang1967}. An eigenstate
$|k\rangle=|k_1,\dots,k_{N+1}\rangle$ is parametrized by $N{+}1$
distinct charge momenta $k_j$ and one rapidity parameter $\Lam$
subject to the Bethe equations
\begin{equation}
\label{eq.bethe}
\ee^{i k_j L}=\frac{\alpha k_j-\Lam+i}{\alpha k_j-\Lam-i}
\qquad\Longleftrightarrow\qquad
\cot\left(\frac{k_jL}{2}\right)=\alpha k_j-\Lam ,
\qquad j=1,\dots,N+1 ,
\end{equation}
supplemented by the one-magnon condition, which quantizes the total
momentum $Q$ and thereby fixes $\Lam$,
\begin{equation}
\label{eq.Q}
\prod_{j=1}^{N+1}\ee^{i k_jL}=1
\qquad\Longrightarrow\qquad
Q=\sum_{j=1}^{N+1}k_j=\frac{2\pi m}{L},\qquad m\in\mathbb{Z} .
\end{equation}
The energy of the state $|k\rangle$ is
\begin{equation}
\label{eq.Ek}
E_k=\frac12\sum_{j=1}^{N+1}k_j^2 .
\end{equation}

Each root $k_j$ lives on a branch labelled by an integer $n_j$,
$2\pi n_j/L<k_j<2\pi(n_j+1)/L$, with
$k_j$ increasing monotonically in $\Lam$,
\begin{equation}
\label{eq.dkdL}
\frac{\partial k_j}{\partial\Lam}
=\frac{2}{L}
\frac{1}{(\alpha k_j-\Lam)^2+1+2\alpha/L}>0 ,
\end{equation}
and $k_j\to 2\pi n_j/L$ as $\Lam\to-\infty$. An eigenstate is therefore
labelled by a set of $N{+}1$ distinct integers $\{n_j\}$ and the total
momentum integer $m$. A solution with finite $\Lam$ exists if
$m-N-1<\sum_j n_j<m$. The boundary case
\begin{equation}
\label{eq.special}
\sum_j n_j=m
\qquad\Longleftrightarrow\qquad
\Lam=-\infty ,\qquad k_j=\frac{2\pi n_j}{L} ,
\end{equation}
is a legitimate eigenstate, the ``special'' state, which we denote
$|k_\infty\rangle$. In it the impurity
is completely decoupled in momentum space, see \eref{eq.special-decomp}
below.

At infinite coupling, $\alpha=0$, the Bethe equations decouple from
$k_j$ and all roots share a common phase shift,
\begin{equation}
\label{eq.alpha0-roots}
\alpha=0:\qquad
k_j=\frac{2\pi n_j+2\delta}{L},
\qquad
\delta=\frac{\pi(m-\textstyle\sum_j n_j)}{N+1}\in(0,\pi),
\qquad
\Lam=-\cot\delta ,
\end{equation}
so that $\Lam$ depends on the branch integers only through their sum:
the spectrum organizes into degenerate-$\Lam$ sectors. This
structure controls the infinite-coupling selection rules.

Let $\hat c^\dagger_q$ create the impurity in the plane-wave state of
momentum $q=2\pi n/L$, and let $|q_1\ldots q_N\rangle$ denote the free
Fermi-gas states, $q_a=2\pi m_a/L$, $m_1<\cdots<m_N$. The central quantity is the overlap formula
\cite{GPZ2015}
\begin{equation}
\label{eq.overlap}
\langle q_1\ldots q_N|\hat c_q|k\rangle
=\delta_{q+\sum_a q_a,\,Q}
\left(\frac{2}{L}\right)^{N/2}
\sqrt{\frac{\prod_{j=1}^{N+1}\frac{\partial k_j}{\partial\Lam}}
{\sum_{j=1}^{N+1}\frac{\partial k_j}{\partial\Lam}}}
\det\begin{pmatrix}
\frac{1}{k_1-q_1}&\cdots&\frac{1}{k_{N+1}-q_1}\\
\vdots&\ddots&\vdots\\
\frac{1}{k_1-q_N}&\cdots&\frac{1}{k_{N+1}-q_N}\\
1&\cdots&1
\end{pmatrix}.
\end{equation}
The bordered determinant in \eref{eq.overlap} evaluates in
closed form (it is the $q_0\to\infty$ limit of a Cauchy determinant with
an auxiliary row $q_0$),
\begin{equation}
\label{eq.overlap-closed}
\det(\cdots)
=(-1)^{\frac{N(N-1)}{2}}
\frac{\prod_{a<a'}(q_{a'}-q_a)\prod_{j<j'}(k_{j'}-k_j)}
{\prod_{a,j}(k_j-q_a)} .
\end{equation}

For the special state, \eref{eq.special}, the overlap formula degenerates
into a free-particle decomposition: taking $\Lam\to-\infty$ in
Eqs.~\eqref{eq.overlap}--\eqref{eq.overlap-closed} gives
\begin{equation}
\label{eq.special-decomp}
\hat c_q|k_\infty\rangle=
\begin{cases}
\dfrac{(-1)^{N+1-s}}{\sqrt{N+1}}
\left|\{2\pi n_j/L\}\setminus q\right\rangle , & q=\frac{2\pi n_s}{L}\in\{k_j\},\\[2mm]
0,&\text{otherwise},
\end{cases}
\end{equation}
i.e. the special state is an equal-weight superposition of the
impurity occupying each of the $N{+}1$ free momenta, the gas filling
the rest.

We study the standard injection (sudden) protocol
\cite{Mathy2012,Knap2014}: at $t=0$ the impurity is created
with momentum $p_0=2\pi m_0/L$ on top of the ground state of the free
Fermi gas, in which the $N$ momenta (odd $N$) fill the Fermi sea,
\begin{equation}
\FS=|q_1\ldots q_N\rangle,\qquad |q_a|\le \kF-\frac{\pi}{L},\qquad
E_{\rm FS}=\frac12\sum_{a=1}^{N}q_a^2 ,
\end{equation}
with $E_{\rm FS}$ its energy. The initial state is
\begin{equation}
\label{eq.psi0}
|\psi(0)\rangle=\hat c^\dagger_{p_0}\FS=\sum_a c_a|k_a\rangle ,
\qquad
c_a=\langle k_a|\hat c^\dagger_{p_0}\FS
\end{equation}
where $|k_a\rangle$, labelled by $a$, runs over all eigenstates with
total momentum $Q=p_0$, and the real overlaps $c_a$ are given by
\eref{eq.overlap}. The state evolves as
$|\psi(t)\rangle=\ee^{-iHt}|\psi(0)\rangle$. The observables of interest
are the average impurity momentum and its variance,
\begin{equation}
\label{eq.Pt-def}
P(t)=\langle\psi(t)|\Pimp|\psi(t)\rangle
=\sum_{a,b}c_ac_b\langle k_b|\Pimp|k_a\rangle
\ee^{i t(E_b-E_a)} ,
\qquad
\Pimp=\sum_q q\hat c_q^\dagger \hat c_q ,
\end{equation}
where $\Pimp$ is the impurity momentum operator, and
\begin{equation}
\label{eq.Var-def}
\operatorname{Var}P(t)=\langle\psi(t)|\Pimp^2|\psi(t)\rangle-P(t)^2 ,
\end{equation}
whose first term has the same spectral form with $\Pimp^2$ in place of
$\Pimp$. Everything thus reduces to the matrix elements
$\langle p|\Pimp^n|k\rangle$.

\section{Matrix elements of the impurity momentum}
\label{sec.matrix-elements}

Between any two eigenstates $|k\rangle$ (roots $k_j$, rapidity $\Lam$)
and $|p\rangle$ (roots $p_i$, rapidity $\Lamt$) with the same total
momentum $Q$, consider the matrix elements $n_q^{(pk)}$ of the
impurity occupation number in momentum $q$ and their generating
function $\chi(y)$, where the displacement $y$ is conjugate to the
impurity momentum,
\begin{equation}
\label{eq.nq-def}
n_q^{(pk)}=\langle p|\hat c^\dagger_q\hat c_q|k\rangle ,
\qquad
\chi(y)=\sum_q \ee^{-i qy}n_q^{(pk)} ,
\qquad
\langle p|\Pimp^{n}|k\rangle=\left.(i\partial_y)^n\chi(y)\right|_{y=0} .
\end{equation}
Inserting the resolution of unity over free states and using
\eref{eq.overlap}, the sum over gas configurations is performed in closed form
(\arf{app:cauchy-binet}), with the result
\begin{equation}
\label{eq.rho-det}
\chi(y)=\det(A+B)-\det A ,
\end{equation}
where $A$ and $B$ are $(N{+}1)\times(N{+}1)$ matrices,
\begin{align}
\label{eq.Adef}
A_{ij}&=-\frac{e(p_i)-e(k_j)}{p_i-k_j}
\ee^{-i y (p_i+k_j)/2}
\left|\frac{\partial p_i}{\partial\Lamt}\right|^{1/2}
\left|\frac{\partial k_j}{\partial\Lam}\right|^{1/2},
\\
\label{eq.Bdef}
B_{ij}&=\frac{\ee^{-i y(p_i+k_j)/2}}
{\sqrt{\sum_i \partial p_i/\partial\Lamt}
\sqrt{\sum_j \partial k_j/\partial\Lam}}
\left|\frac{\partial p_i}{\partial\Lamt}\right|^{1/2}
\left|\frac{\partial k_j}{\partial\Lam}\right|^{1/2},
\end{align}
and, for $0<y<L$,
\begin{equation}
\label{eq.efun}
e(\kappa)=\frac2L\sum_{n\in\mathbb{Z}}
\frac{\ee^{2\pi i ny/L}}{\kappa-2\pi n/L}
=\ee^{i\kappa y}\left[\cot\left(\frac{\kappa L}{2}\right)-i\right]
\xrightarrow{\ \text{on-shell}\ }
\ee^{i\kappa y}(\alpha\kappa-\Lam_\kappa-i) ,
\end{equation}
where ``on-shell'' means that $\kappa$ is a Bethe root, $\Lam_\kappa$
is the rapidity of the state it belongs to, and the Bethe equations
\eqref{eq.bethe} have been used. When the two states share a root, $p_i=k_j=\kappa$ (possible only
in the degenerate sectors at $\alpha=0$), the corresponding entry of
$A$ differs from the naive limit of \eref{eq.Adef}. The underlying
lattice sum gives
\begin{equation}
\label{eq.shared-entry}
\left.A_{ij}\right|_{p_i=k_j=\kappa}
=\left[i y(\alpha\kappa-\Lam-i)
-\frac{L}{2}\left(1+(\alpha\kappa-\Lam)^2\right)\right]
\frac{\partial k_j}{\partial\Lam}
=\left(\text{naive limit}\right)-1 ,
\end{equation}
where the naive limit is the $p_i\to k_j$ limit of \eref{eq.Adef} with
$e(\kappa)$ in its on-shell form, and we used \eref{eq.dkdL}. This shift by $-1$ is the origin of
the selection rules. Equations
\eqref{eq.rho-det}--\eqref{eq.shared-entry} hold in finite volume.

Expanding \eref{eq.rho-det} to second order in $y$
(\arf{app:smally}) one finds for the moments of the diagonal elements
($p=k$) \cite{GLZ2020}
\begin{equation}
\label{eq.P-diag}
\langle k|\Pimp|k\rangle=\frac{S_1}{S_0},
\qquad
\langle k|\Pimp^2|k\rangle=\frac{S_2-\alpha(S_0S_2-S_1^2)}{S_0},
\qquad
S_n\equiv\sum_{j=1}^{N+1}k_j^n
\frac{\partial k_j}{\partial\Lam} ,
\end{equation}
while for $\Lamt\neq\Lam$ the off-diagonal elements are
\begin{align}
\label{eq.P-offdiag}
\langle p|\Pimp|k\rangle&=
\frac{D(k,p)}
{\sqrt{\sum_i\frac{\partial p_i}{\partial\Lamt}}
\sqrt{\sum_i\frac{\partial k_i}{\partial\Lam}}}
\frac{E_p-E_k}{\Lamt-\Lam} ,
\\[1mm]
\label{eq.P2-offdiag}
\langle p|\Pimp^2|k\rangle&=
\frac{D(k,p)}
{\sqrt{\sum_i\frac{\partial p_i}{\partial\Lamt}}
\sqrt{\sum_i\frac{\partial k_i}{\partial\Lam}}}
\left[\frac{\sum_i p_i^3-\sum_i k_i^3}{3(\Lamt-\Lam)}
+\alpha\left(\frac{E_p-E_k}{\Lamt-\Lam}\right)^{2}\right] ,
\end{align}
with the energies $E_p$ and $E_k$ of \eref{eq.Ek} and the
determinant
\begin{equation}
\label{eq.Ddet}
D(k,p)=\det\left[\frac{\Lamt-\Lam}{p_i-k_j}
\left|\frac{\partial p_i}{\partial\Lamt}\right|^{1/2}
\left|\frac{\partial k_j}{\partial\Lam}\right|^{1/2}\right]
=(\Lamt-\Lam)^{N+1}\;
\frac{\prod\limits_{i<j}(p_j-p_i)(k_i-k_j)}
{\prod\limits_{i,j}(p_i-k_j)}
\prod_{j=1}^{N+1}
\left|\frac{\partial p_j}{\partial\Lamt}
\frac{\partial k_j}{\partial\Lam}\right|^{1/2} .
\end{equation}
Equations \eqref{eq.P-diag}--\eqref{eq.Ddet} are the first main result
of this paper. Their derivation combines the rank-one structure of the
$y$-expansion of \eref{eq.rho-det} with contour-integral
representations of the inverse of the matrix
$C_{ij}=(\Lamt-\Lam)/(p_i-k_j)$, whose summed entries reduce to power
sums of the roots, \eref{eq.Cinv-sums} in \arf{app:smally}.

The off-diagonal elements satisfy
$\langle p|\Pimp|k\rangle\propto E_p-E_k$: the impurity momentum
operator connects only states of different energy (for
$\Lamt\ne\Lam$), a structure reminiscent of an operator that is a
total time derivative, here of the impurity position operator
weighted by the mass ratio. The sum
$\sum_qq^2n_q^{(pk)}$ defining $\langle p|\Pimp^2|k\rangle$ is only
conditionally convergent: $n_q^{(pk)}$ has a $O(q^{-3})$ tail with
alternating sign (the diagonal tail is the Tan-contact decay
$\propto q^{-4}$), and \eref{eq.P2-offdiag} resums it in the
symmetric (equivalently Abel) sense, which is also the sense in which
the generating function \eref{eq.nq-def} is differentiated.

Matrix elements involving the special state \eref{eq.special} follow
from the $\Lam\to-\infty$ limit (equivalently from
\eref{eq.special-decomp}). The diagonal elements are
\begin{equation}
\label{eq.special-diag}
\langle k_\infty|\Pimp|k_\infty\rangle=\frac{Q}{N+1},
\qquad
\langle k_\infty|\Pimp^2|k_\infty\rangle=\frac{2E_k}{N+1} ,
\end{equation}
and for a generic partner $|p\rangle$,
\begin{equation}
\label{eq.special-offdiag}
\begin{aligned}
\langle p|\Pimp|k_\infty\rangle
&=\frac{D_\infty(k,p)}
{\sqrt{\sum_i\frac{\partial p_i}{\partial\Lamt}}\sqrt{2(N{+}1)/L}}
\sum_{j=1}^{N+1}\frac{p_j^2-k_j^2}{2},\\
\langle p|\Pimp^2|k_\infty\rangle
&=\frac{D_\infty(k,p)}
{\sqrt{\sum_i\frac{\partial p_i}{\partial\Lamt}}\sqrt{2(N{+}1)/L}}
\sum_{j=1}^{N+1}\frac{p_j^3-k_j^3}{3},
\end{aligned}
\end{equation}
\begin{equation}
\label{eq.Dinf}
D_\infty(k,p)=\det\left[
\frac{2/L}{(p_i-k_j)\sqrt{(\alpha p_i-\Lamt)^2+1+2\alpha/L}}
\right] ,
\end{equation}
where the $k_j=2\pi n_j/L$ are the free roots of the special state.
Matrix elements between two distinct special states vanish
for all $n$, as is evident from
\eref{eq.special-decomp}: their $\hat c_q$-decompositions share no
common free state.

For one gas particle and total momentum $Q=0$ the eigenstates come in
two families: special states $|q\rangle\equiv|q,-q\rangle$ with
$q\in\frac{2\pi}{L}\mathbb{N}$ and $\Lam=-\infty$, and generic states
$|k\rangle=|k,-k\rangle$ with $\Lam=0$, where $\cot(kL/2)=\alpha k$,
$k>0$. Our formulas give the only nonvanishing matrix elements of
$\Pimp$ as
\begin{equation}
\label{eq.N1-P}
\langle q|\Pimp|k\rangle=\frac{2}{L}
\frac{2kq}{\sqrt{(\alpha k)^2+1+2\alpha/L}}
\frac{1}{E_q-E_k},
\qquad E_q=q^2,E_k=k^2 ,
\end{equation}
and, for $\Pimp^2$,
\begin{equation}
\label{eq.N1-P2}
\langle k'|\Pimp^2|k\rangle=-\frac{2\alpha}{L}
\frac{2kk'}
{\sqrt{(\alpha k)^2{+}1{+}\frac{2\alpha}{L}}
\sqrt{(\alpha k')^2{+}1{+}\frac{2\alpha}{L}}},
\qquad
\langle k|\Pimp^2|k\rangle=k^2
\frac{1+(\alpha k)^2-2\alpha/L}{1+(\alpha k)^2+2\alpha/L} .
\end{equation}
As a check, consider the operator identity
\begin{equation}
\Pimp^2=\sum_q\Pimp|q\rangle\langle q|\Pimp .
\end{equation}
Performing the sum over
the special states with \eref{eq.N1-P} and the partial-fraction
summation formula
\begin{equation}
\sum_{q>0}\frac{E_q}{(E_q-E_k)(E_q-E_{k'})}
=\frac{L\left[k'\cot(k'L/2)-k\cot(kL/2)\right]}{4(k^2-k'^2)} ,
\end{equation}
together with the Bethe equations, reproduces both formulas in
\eref{eq.N1-P2}.

In the degenerate-$\Lam$ sectors of \eref{eq.alpha0-roots}, formulas
\eqref{eq.P-offdiag}--\eqref{eq.Ddet} degenerate into $0/0$. The
resolution is the shared-root entry \eref{eq.shared-entry}. Two
distinct states of the same sector necessarily differ by $u\ge2$ roots
(one displaced root would change $\sum_jn_j$), and ordering the shared
roots first one finds the small-$y$ structure
\begin{equation}
\label{eq.deg-structure}
A(y)=-\Delta_r+i ycJ+O(y^3),
\qquad
B(y)=\frac{b\otimes\tilde b}{N+1}\ \ \text{(rank one, all $y$)},
\end{equation}
where $\Delta_r$ is the projector on the $r=N{+}1-u$ shared slots,
$J$ is the all-ones matrix, and
\begin{equation}
c=-(\Lam+i)\frac{\partial k}{\partial\Lam},\qquad
b_i=\ee^{-i yp_i/2},\qquad
\tilde b_j=\ee^{-i yk_j/2} .
\end{equation}
Expanding
$\det(A+B)-\det A$ by multilinearity, at order $y$ every term contains
two columns proportional to the same vector, and at order $y^2$
(for $u=2$) the only two surviving border terms cancel pairwise.
For $u\ge3$ the expansion starts at $O(y^3)$ (\arf{app:degenerate}). This
gives the infinite-coupling selection rule: at $\alpha=0$, for any two
distinct eigenstates with equal rapidity $\Lam$ (equal $\sum_j n_j$,
including $\Lambda=-\infty$),
\begin{equation}
\label{eq.selection}
\langle p|\Pimp|k\rangle=0
\qquad\text{and}\qquad
\langle p|\Pimp^2|k\rangle=0 .
\end{equation}

The second infinite-coupling statement concerns the diagonal:
since at $\alpha=0$ all $\partial k_j/\partial\Lam$ in
\eref{eq.P-diag} are equal, every eigenstate (special ones included)
carries
\begin{equation}
\label{eq.P-diag-inf}
\alpha=0:\qquad
\langle k|\Pimp|k\rangle=\frac{1}{N+1}\sum_{j=1}^{N+1}k_j
=\frac{Q}{N+1},
\qquad
\langle k|\Pimp^2|k\rangle=\frac{2E_k}{N+1} .
\end{equation}
At infinite repulsion the impurity, which has the mass of a host
fermion, carries its per-particle share of the total momentum and
kinetic energy in every eigenstate. Both statements fail at any finite coupling.

\section{Momentum relaxation}
\label{sec.dynamics}

With the closed matrix elements of \srf{sec.matrix-elements}, the
spectral sum \eref{eq.Pt-def} becomes an explicit representation of
the momentum dynamics, which we evaluate in \srf{sec.numerics}. Before
doing so, we extract the stationary value.

In a finite system $P(t)$ is quasiperiodic and its stationary value is
understood in the Ces\`aro sense,
\begin{equation}
P_\infty\equiv\lim_{T\to\infty}\frac1T\int_0^TP(t)\,d t .
\end{equation}
Naively, $P_\infty$ equals the diagonal-ensemble average, given by the
diagonal ($a=b$) terms of \eref{eq.Pt-def}, up to contributions of accidentally degenerate pairs $E_a=E_b$, $a\ne b$. For the impurity
momentum these contributions vanish: every
off-diagonal matrix element of $\Pimp$ in
Eqs.~\eqref{eq.P-offdiag}, \eqref{eq.special-offdiag} is proportional
to $E_p-E_k$, matrix elements between distinct special states vanish,
and within degenerate-$\Lam$ sectors at $\alpha=0$ they vanish by
the selection rule \eref{eq.selection}. Hence
\begin{equation}
\label{eq.Pinf-diag}
P_\infty=\sum_a c_a^2\langle k_a|\Pimp|k_a\rangle
=\sum_a c_a^2\frac{S_1^{(a)}}{S_0^{(a)}} ,
\end{equation}
including degenerate pairs, where $S_n^{(a)}$ is the sum $S_n$ of
\eref{eq.P-diag} evaluated on $|k_a\rangle$. For $\Pimp^2$ the analogous statement
holds up to the cubic terms in \eref{eq.P2-offdiag}, which do not
vanish at $E_p=E_k$.

At infinite coupling \eref{eq.P-diag-inf} turns
\eref{eq.Pinf-diag} into a truncation-free statement. Since
every eigenstate in the sector $Q=p_0$ carries impurity momentum
$p_0/(N{+}1)$ and $\sum_ac_a^2=1$,
\begin{equation}
\label{eq.Pinf-inf}
\gamma=\infty:\qquad
P_\infty=\frac{p_0}{N+1}\xrightarrow[\ N\to\infty\ ]{}0 .
\end{equation}
At infinite repulsion the mean momentum of the equal-mass impurity
relaxes completely, for any initial momentum $p_0$: the
incomplete-relaxation plateau
\cite{Mathy2012,Lychkovskiy2015} closes as
$\gamma\to\infty$. This agrees with Ref.~\cite{GLZ2018}, where the
stationary impurity velocity after injection was obtained in the
thermodynamic limit and found to vanish as $\gamma\to\infty$.

At finite coupling $S_1/S_0$ varies from state to state and
$P_\infty>0$. In the thermodynamic limit the sum \eref{eq.Pinf-diag}
becomes an integral over the rapidity,
\begin{equation}
\label{eq.Pinf-W}
P_\infty=\int d\Lam\,\mathcal W(\Lam)P_{\rm pol}(\Lam) ,
\end{equation}
where $P_{\rm pol}(\Lam)=S_1/S_0$ is the polaron momentum of the
$\Lam$-family, \eref{eq.P-diag}, and $\mathcal W(\Lam)$ is the rapidity-resolved
overlap weight of the injected state, which for the injection protocol
is its stationary rapidity distribution. We evaluate
\eref{eq.Pinf-diag} directly from the finite-size sums in
\srf{sec.numerics}.

A second statement at $\gamma=\infty$ concerns the variance.
By \eref{eq.P-diag-inf} the diagonal-ensemble value of $\Pimp^2$ is
proportional to the mean energy of the injected state,
\begin{equation}
\langle\Pimp^2\rangle_\infty=\frac{2}{N+1}\sum_ac_a^2E_a
=\frac{2}{N+1}\langle\psi(0)|H|\psi(0)\rangle ,
\end{equation}
and this energy,
\begin{equation}
\label{eq.energy-sumrule}
\langle\psi(0)|H|\psi(0)\rangle
=\frac{p_0^2}{2}+E_{\rm FS}+g\rho ,
\end{equation}
diverges linearly with $g$. Hence at strictly infinite coupling the
momentum variance of the injected impurity never saturates: the
overlap weight $c_a^2$ of eigenstates in which a particle is excited
to a large momentum $\kappa$ decays only as $\kappa^{-2}$ (the
initial-state analogue of the Tan contact), and this tail feeds the variance
indefinitely, and $\operatorname{Var}P(t)$ grows without bound, while
at any finite $\gamma$ it saturates at the (finite) diagonal-ensemble
value.

The spectral sum \eref{eq.Pt-def} holds in finite volume. We now
recast it, in the thermodynamic limit, as a rapidity integral over
Fredholm determinants, the analogue for $P(t)$ of the known
determinant representations of the impurity Green's function.
The derivation is given in Apps.~\ref{app:pt} and \ref{app:alpha0}.
Here we describe
the structure and the result.

The double spectral sum is organized by the two rapidities, now
written $\Lam_k$ and $\Lam_p$ for the states $|k\rangle$ and
$|p\rangle$ (the $\Lam$ and $\Lamt$ of \srf{sec.matrix-elements}). The
sum over eigenstates $|k\rangle$ ($|p\rangle$) is traded for an
integral over $\Lam_k$ ($\Lam_p$) and a free sum over branch
configurations,
\begin{equation}
\label{eq.sum-to-int}
\sum_{k}
=\int d\Lam_k
\sum_{n_1<\cdots<n_{N+1}}\int\frac{d x}{2\pi}
\left(\sum_{j=1}^{N+1}\frac{\partial k_j}{\partial\Lam_k}\right)
\ee^{i x\left(\sum_j k_j-p_0-\sum_j q_j\right)} ,
\end{equation}
where the auxiliary $x$-integral enforces momentum conservation
relative to the momenta $\{q_j\}$ of $\FS$; the sum over $|p\rangle$
carries the analogous auxiliary variable $y$. Using the overlap formula
\eref{eq.overlap}, the matrix element \eref{eq.P-offdiag}, and the
subset summation of \arf{app:cauchy-binet}, the sums over
both branch configurations collapse into a single
$(N{+}1)\times(N{+}1)$ bordered determinant whose entries are double
sums over Bethe branches. Energy denominators are removed by computing
the primitive $\Phi(t)$, defined by $P(t)=-i\,d\Phi/d t$
(\arf{app:pt}). In the thermodynamic limit the branch sums turn into
contour integrals with simple closed-form limits, and the bordered
determinant becomes a Fredholm determinant on the Fermi sea
$[-\kF,\kF]$. The final result reads
\begin{equation}
\label{eq.Pt-final}
\begin{aligned}
P(t)={}&\Re \iint d\Lam_k\,d\Lam_p
\iint\frac{d x\,d y}{(2\pi)^2}
\ee^{i(y-x)p_0}\\
&\times\left[\vphantom{\Et_0\det(1+\hat K)}
\Et_0p_0\left((E_0-1)\det(1+\hat K)+\det(1+\hat K-\hat K_1)\right)\right.\\
&\qquad+\det(1+\hat K+\hat K_1+\hat K_2)-\det(1+\hat K+\hat K_1)\\
&\qquad-\left.\det(1+\hat K+\hat K_2)
+\det(1+\hat K-E_0\hat K_2+\Gamma\hat K_3)\vphantom{\Et_0\det(1+\hat K)}\right] ,
\end{aligned}
\end{equation}
where $\Re$ denotes the real part, the hatted symbols are integral
operators acting on functions on $[-\kF,\kF]$, $\det$ is their Fredholm
determinant, and their kernels are
\begin{equation}
\label{eq.kernels}
\begin{gathered}
K(q,q')=\frac{\Lam_k-\Lam_p}{\pi}
\frac{v(q)\ee^{-i q'x}-\ee^{i qy}w(q')+\Et(q)E(q')}{q-q'} ,\\
K_1(q,q')=\frac{\Lam_k-\Lam_p}{\pi}v(q)E(q'),\qquad
K_2(q,q')=\frac{\Lam_k-\Lam_p}{\pi}\tilde h(q)w(q'),\\
K_3(q,q')=\frac{\Lam_k-\Lam_p}{\pi}\tilde h(q)E(q'),\qquad
\tilde h(q)=q\Et(q)+2\Et_0\ee^{i qy} .
\end{gathered}
\end{equation}
The scalar functions
$v$, $w$, $E$, $\Et$, $E_0$, $\Et_0$ and $\Gamma$ are elementary contour integrals listed
in \arf{app:pt} [Eqs.~\eqref{eq.v-int}--\eqref{eq.E0-int}]; they
depend on $(x,y,t;\Lam_k,\Lam_p)$ and are smooth except for
principal-value and $\delta$-type distributions at coinciding
arguments. The kernel $K$ is regular at $q=q'$ by virtue of the
identity
\begin{equation}
\label{eq.regularity}
v(q)\ee^{-i qx}-\ee^{i qy}w(q)+\Et(q)E(q)=0 ,
\end{equation}
which at infinite coupling takes the explicit form \eref{eq.identity98}
below. The representation \eref{eq.Pt-final}
is manifestly nonsingular at $\Lam_k=\Lam_p$. A representation for
the primitive $\Phi(t)$, involving the same kernels but only two
determinants, is also available [\arf{app:pt},
\eref{eq.Bt-final}] and can be preferable when the integration by
parts leading to \eref{eq.Pt-final} requires justification.

At $\alpha=0$ every contour integral entering the kernels evaluates in
closed form through the Fresnel error function (\arf{app:alpha0}):
\begin{align}
\label{eq.v-erf}
v(q)&=\frac{\ee^{i xq}}{(1{+}\Lam_k^2)(1{+}\Lam_p^2)}
\left[(\Lam_k\Lam_p{+}1)
\frac{\Lam_p-i\sgn(x{-}y)}{\Lam_k-\Lam_p}
+i\Lam_p\Erf\left(\frac{(x-qt)(1-i)}{2\sqrt t}\right)
+i V(q)\right] ,\\
\label{eq.w-erf}
w(q)&=\frac{\ee^{-i yq}}{(1{+}\Lam_k^2)(1{+}\Lam_p^2)}
\left[(\Lam_k\Lam_p{+}1)
\frac{\Lam_k-i\sgn(x{-}y)}{\Lam_k-\Lam_p}
+i\Lam_k\Erf\left(\frac{(y-qt)(1+i)}{2\sqrt t}\right)
+i W(q)\right] ,
\end{align}
\begin{align}
\label{eq.VW-erf}
V(q)&=\frac2\pi\int_{|x-y|/t}^{\infty}d k\,
\frac{\ee^{-i tk^2/2+i k\sgn(x-y)(x-qt)}}{k}
-i\sgn(x{-}y)\Erf\left(\frac{(x-qt)(1-i)}{2\sqrt t}\right) ,\\
W(q)&=\frac2\pi\int_{|x-y|/t}^{\infty}d k\,
\frac{\ee^{i tk^2/2+i k\sgn(x-y)(y-qt)}}{k}
-i\sgn(x{-}y)\Erf\left(\frac{(y-qt)(1+i)}{2\sqrt t}\right) ,
\end{align}
\begin{equation}
\label{eq.Et-erf}
\begin{gathered}
E(q)=\frac{i\ee^{-i tq^2/2}}{\Lam_k^2+1}
\Erf\left(\frac{(x-qt)(1-i)}{2\sqrt t}\right)
-\frac{\Lam_k\ee^{-i tq^2/2}}{\Lam_k^2+1},\\
\Et(q)=-\frac{i\ee^{i tq^2/2}}{\Lam_p^2+1}
\Erf\left(\frac{(y-qt)(1+i)}{2\sqrt t}\right)
-\frac{\Lam_p\ee^{i tq^2/2}}{\Lam_p^2+1},
\end{gathered}
\end{equation}
\begin{equation}
\label{eq.gauss-factors}
\begin{gathered}
E_0=\sqrt{\frac{2}{i\pi t}}
\frac{\ee^{i x^2/(2t)}}{1+\Lam_k^2},
\qquad
\Et_0=\sqrt{\frac{2i}{\pi t}}
\frac{\ee^{-i y^2/(2t)}}{1+\Lam_p^2},\\
\Gamma=\frac{2\ee^{i\frac{x^2-y^2}{2t}}}
{(\Lam_p^2{+}1)(\Lam_k^2{+}1)}
\left[\frac{(1{+}\Lam_p\Lam_k)\delta(y{-}x)}{\Lam_p-\Lam_k}
+\frac{1}{\pi(y-x)}\right] .
\end{gathered}
\end{equation}
The regularity of the kernel at $q=q'$ is now the explicit identity
\begin{equation}
\label{eq.identity98}
i\left[V(q)-W(q)\right]
+\Erf\left(\frac{(x-qt)(1-i)}{2\sqrt t}\right)
\Erf\left(\frac{(y-qt)(1+i)}{2\sqrt t}\right)-1=0 ,
\end{equation}
built on the principal-value integral \eref{eq.vp-erf}.

At $t=0$ the Gaussian factors become $\delta$-functions and the
kernels lose their $t$-dependence, \eref{eq.t0-kernels} in
\arf{app:alpha0}. The
term $\Et_0p_0E_0\det(1+\hat K)$ in \eref{eq.Pt-final} then
collapses the $x$ and $y$ integrations to $x=y=0$, where
$\hat K$ vanishes. Using $\int d\Lam\,(1+\Lam^2)^{-1}=\pi$ one finds
that this single term already saturates the answer,
\begin{equation}
\label{eq.P0-emergence}
\Re\iint d\Lam_kd\Lam_p
\iint\frac{d xd y}{(2\pi)^2}\ee^{i(y-x)p_0}
\Et_0p_0E_0\det(1+\hat K)
=p_0\frac{1}{\pi^2}\iint
\frac{d\Lam_k\,d\Lam_p}{(1{+}\Lam_k^2)(1{+}\Lam_p^2)}
=p_0 ,
\end{equation}
so that consistency, $P(0)=p_0$, requires the remaining terms to
cancel after the $(x,y,\Lam_k,\Lam_p)$ integrations. In the
finite-size spectral representation the cancellation is automatic due to the
completeness of the eigenbasis.

At $t\to\infty$ with $q,x,y=O(1)$ the error functions in
Eqs.~\eqref{eq.v-erf}--\eqref{eq.Et-erf} saturate to $-\sgn(q)$, and
the Gaussian factors decay as $t^{-1/2}$. The functions
$v,w,E,\Et$ approach the $t$-independent limits given in
\arf{app:alpha0} [Eqs.~\eqref{eq.tinf-limits}], so the explicitly
$p_0$-dependent first line of \eref{eq.Pt-final} dies out as
$t^{-1/2}$ and the stationary value is carried entirely by the
four-determinant combination in the second and third lines. Its value
must reproduce the diagonal-ensemble result, which at infinite
coupling is $P_\infty=0$ in the thermodynamic limit,
\eref{eq.Pinf-inf}. 

\section{Relaxation dynamics in finite systems}
\label{sec.numerics}

We evaluate the spectral sums \eref{eq.Pt-def} and
\eref{eq.Pinf-diag} in finite systems from the product forms
\eqref{eq.overlap-closed} and \eqref{eq.Ddet}, enumerating eigenstates by their
particle--hole content relative to the Fermi sea and retaining those
with the largest overlap weight $c_a^2$. The truncation is monitored by
the sum rules for the norm, for the initial momentum $P(0)=p_0$ and, at
finite coupling, for the energy \eref{eq.energy-sumrule}. At
$\gamma=\infty$ every state missing from the retained basis carries
diagonal momentum $p_0/(N{+}1)$, \eref{eq.P-diag-inf}, so the
stationary value does not depend on the truncation. Times are measured
in units of the Fermi time $1/\EF$. We call the injection subsonic
when $p_0<\kF$, so that the initial impurity velocity is below the
Fermi velocity, and supersonic when $p_0>\kF$. Finite-size dynamics represents the thermodynamic
limit up to the traversal time $t^*\simeq L/\vF$, after which ring
revivals appear. Comparing several system sizes at fixed $p_0/\kF$
exposes this window.

\begin{figure}[t]
\centering
\includegraphics[width=\textwidth]{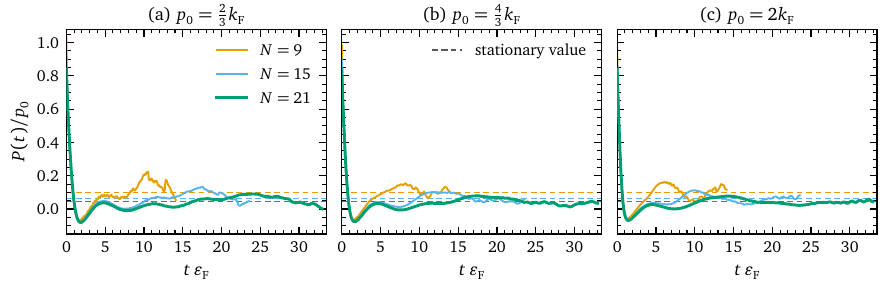}
\caption{Momentum relaxation at infinite coupling for
$p_0/\kF=\frac23,\frac43,2$ and $N=9,15,21$ (color; heavier line
$N=21$). Time
axes end at the traversal time $t^*$ of each system. Dashed lines:
the stationary values \eref{eq.Pinf-inf}, which vanish in the thermodynamic limit. Curves at different $N$
collapse onto a common thermodynamic-limit envelope for
$t\lesssim t^*$.}
\label{fig.relax-inf}
\end{figure}

Figure~\ref{fig.relax-inf} shows $P(t)/p_0$ at $\gamma=\infty$.
The mean momentum collapses on the scale of a Fermi time, for subsonic
and supersonic injection alike. The relaxation then overshoots: $P(t)$
turns negative, so that transiently the impurity moves, on average,
opposite to its injection direction, a backscattering echo of the
Fermi sea. The curves then settle onto the plateau
\eref{eq.Pinf-inf} (dashed lines). For subsonic
injection the accessible window retains slow oscillations around the
plateau whose position scales with $t^*$ and whose amplitude decreases
with $N$, the signature of ring revivals. Oscillations of
thermodynamic-limit origin around the plateau are weak at
$\gamma=\infty$: pronounced flutter is a finite-coupling phenomenon,
see below.

\begin{figure}[t]
\centering
\includegraphics[width=\textwidth]{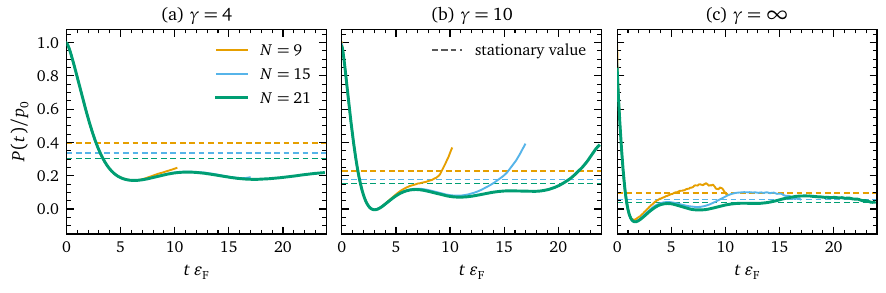}
\caption{Momentum relaxation at $p_0=\frac43\kF$ for
$\gamma=4,10,\infty$ and $N=9,15,21$. Dashed lines: the
diagonal-ensemble stationary values \eref{eq.Pinf-diag} evaluated
from the same spectral data. At finite coupling the relaxation is
incomplete and accompanied by quantum-flutter oscillations.}
\label{fig.relax-gamma}
\end{figure}

Figure~\ref{fig.relax-gamma} contrasts couplings at fixed $p_0$. At
finite coupling the stationary
momentum is a substantial fraction of $p_0$, and it grows as the
coupling decreases. The initial collapse is followed by a slow
quantum-flutter oscillation \cite{Mathy2012,Knap2014,Dolgirev2021}
whose period, within the resolution of the accessible window, is the
same at the two finite couplings shown, echoing the coupling
robustness observed in Ref.~\cite{Knap2014}.
The approach to the diagonal-ensemble plateau \eref{eq.Pinf-diag} is
correspondingly slow at finite coupling: within the window the
oscillation is still centred below $P_\infty$, which the dephasing
argument identifies as the $t\to\infty$ limit
of the finite system.

\begin{figure}[t]
\centering
\includegraphics[width=0.62\textwidth]{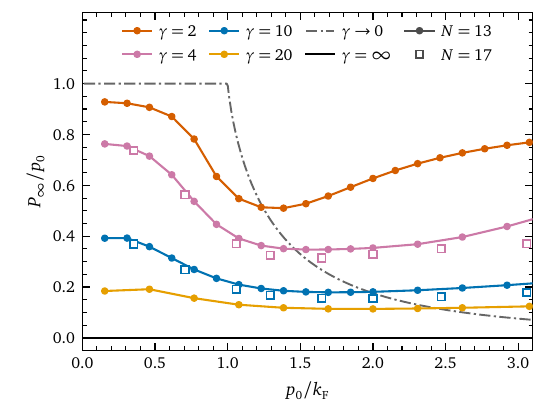}
\caption{Stationary momentum after injection,
\eref{eq.Pinf-diag}, versus injection momentum for several couplings
($N=13$ lines, $N=17$ open squares; the small downward drift of the
squares indicates the residual finite-size effect). Dash-dotted: the
weak-coupling limit \cite{Burovski2014}, equal to $p_0$ below
$\kF$; solid black: the infinite-coupling value zero,
\eref{eq.Pinf-inf}.}
\label{fig.pinf}
\end{figure}

Figure~\ref{fig.pinf} maps $P_\infty/p_0$ over the
$(p_0,\gamma)$ plane using the diagonal ensemble
\eref{eq.Pinf-diag}, whose evaluation requires only overlaps and the
diagonal elements \eref{eq.P-diag}. Below
$\kF$, $P_\infty/p_0$ interpolates monotonically between the two
limits: complete momentum retention at weak coupling (dash-dotted
curve) and complete relaxation
\eref{eq.Pinf-inf} at $\gamma=\infty$. Above $\kF$ the dependence on
the coupling is non-monotonic: each finite-$\gamma$ curve
develops a minimum above $\kF$ and then rises again
(momentum retention grows once the impurity outruns the efficient
particle--hole emission window), while the weak-coupling curve
drops steeply, crosses the moderate-coupling
data and falls below them near $p_0=2\kF$. The kink developing at $p_0=\kF$
separates the subsonic polaron regime from the supersonic emission
regime. We note that the finite-size map of $P_\infty$ is the momentum-space
counterpart of the thermodynamic-limit stationary velocity discussed
in \srf{sec.dynamics}.

\begin{figure}[t]
\centering
\includegraphics[width=0.8\textwidth]{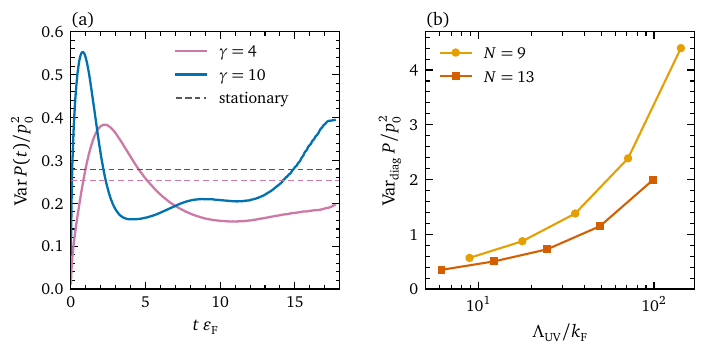}
\caption{Left: variance of the impurity momentum at finite coupling
($N=15$, $p_0=\frac43\kF$); dashed lines are the diagonal-ensemble
stationary values. Right: at $\gamma=\infty$ the stationary
(diagonal-ensemble) variance
$\frac{2}{N+1}\sum_ac_a^2E_a-P_\infty^2$ never converges as the
momentum cutoff $\Lambda_{\rm UV}$ of the state enumeration grows;
the logarithmic axis makes the unbounded (asymptotically linear in
$\Lambda_{\rm UV}$) growth evident.}
\label{fig.variance}
\end{figure}

Finally, \frf{fig.variance} shows the variance \eref{eq.Var-def}. At finite
coupling the variance rises from zero (the initial state is a
momentum eigenstate) and settles into oscillation about the
diagonal-ensemble value (dashed) on the same time scale as the mean.
At $\gamma=\infty$ the diagonal-ensemble variance grows linearly with
the momentum cutoff (right panel), as anticipated in
\srf{sec.dynamics}. The divergence is a property of the idealized
sudden injection at strictly infinite coupling, while at large finite
$\gamma$ the variance saturates at a value $\propto g$,
\eref{eq.energy-sumrule}.

\section{Conclusions and outlook}
\label{sec.conclusions}

We have obtained closed determinant formulas for the matrix elements of
$\Pimp$ and $\Pimp^2$ between arbitrary eigenstates of the McGuire
model, Eqs.~\eqref{eq.P-diag}--\eqref{eq.Ddet} and
\eqref{eq.special-diag}--\eqref{eq.Dinf}. They reduce the momentum
dynamics of an injected impurity to a few determinants per pair of
eigenstates, and they expose structural statements that
had not been noticed: all off-diagonal elements of $\Pimp$ are
proportional to $E_p-E_k$. At infinite coupling both $\Pimp$ and
$\Pimp^2$ obey the degenerate-sector selection rule
\eref{eq.selection}, and at infinite coupling every
eigenstate carries the per-particle momentum share $Q/(N{+}1)$, so
that the injected impurity relaxes completely, \eref{eq.Pinf-inf},
and the incomplete-relaxation plateau
closes at $\gamma=\infty$. The finite-size dynamics shows how this
happens in real time: at $\gamma=\infty$ the momentum collapses within
about one Fermi time, overshoots to negative values, and settles on the
$1/(N{+}1)$ plateau, while at finite coupling the relaxation stops at a
finite $P_\infty$, approached through quantum-flutter oscillations.
In the thermodynamic limit we derived a Fredholm-determinant
representation for $P(t)$, \eref{eq.Pt-final}, with elementary
error-function kernels at $\gamma=\infty$, and analysed its $t=0$ and
$t\to\infty$ limits.

Several problems remain open. (i) The long-time asymptotic
analysis of \eref{eq.Pt-final}, which would give the flutter
frequency and the decay law of $P(t)-P_\infty$ analytically, is a
well-posed steepest-descent/Riemann--Hilbert problem, structurally
similar to the analysis of the impurity Green's function: the $t^{-1/2}$ decay of the explicit
$p_0$-dependent terms is already manifest. (ii) The generating
function \eref{eq.rho-det} gives the matrix elements of
$\ee^{-i y\Pimp}$-like insertions at all orders in $y$, and with them
the full time-dependent momentum distribution $n_q(t)$. (iii) The variance growth at infinite coupling, rooted in the
divergent injection energy \eref{eq.energy-sumrule}, deserves a
quantitative treatment, including the crossover time scale at large
but finite $\gamma$. (iv) Extensions to finite temperature (along the
lines of Refs.~\cite{GPS2022,GPS2023}), to the attractive
branch, to kicked initial states, and to a
Tonks--Girardeau bosonic host are straightforward at the level of the
overlaps and should be within reach. Finally, the infinite-coupling
statements, complete relaxation of the mean momentum with per-particle
momentum sharing, provide a parameter-free benchmark for cold-atom experiments now able to track
impurity momenta in the strongly interacting regime
\cite{Meinert2017,Horvath2026}.

\begin{appendix}
\numberwithin{equation}{section}


\section{Cauchy--Binet reduction of the generating function}
\label{app:cauchy-binet}

Here we derive the determinant representation
\eqref{eq.rho-det}--\eqref{eq.shared-entry} of the generating function
$\chi(y)$ of \eref{eq.nq-def}. Inserting the resolution of
unity over the free basis in \eref{eq.nq-def} and using the overlap
formula \eref{eq.overlap} for both states (with rapidities
$\Lamt,\Lam$; $O_p$ and $O_k$ denote the factors multiplying the
determinant in \eref{eq.overlap} for $|p\rangle$ and $|k\rangle$),
\begin{equation}
\label{eq.rho-sum}
\chi(y)=O_pO_k\sum_{m_1<\cdots<m_N}
\ee^{-i y\left(Q-\sum_a q_a\right)}
\det M^{(p)}\det M^{(k)} ,
\qquad q_a=\frac{2\pi m_a}{L},
\end{equation}
where $M^{(k)}$ is the $(N{+}1)\times(N{+}1)$ matrix of
\eref{eq.overlap} and the sum runs over all $N$-subsets of
$\frac{2\pi}{L}\mathbb{Z}$ (the impurity momentum is fixed by momentum
conservation and produces the factor $\ee^{-i yQ}\ee^{i y\sum_aq_a}$).

The row of ones is generated as a limit: appending an auxiliary
momentum $q_0$ and using
\begin{equation}
\lim_{q_0\to\infty}\frac{-q_0}{k_j-q_0}=1 ,
\end{equation}
each bordered determinant is
the limit of an ordinary determinant with the extended set
$\{q_0,q_1,\dots,q_N\}$. The sum over subsets of the products of two
determinants sharing the index set $\{q_a\}$ is then performed by the
Cauchy--Binet theorem: writing
$X^{(k)}_{n,j}=1/(k_j-2\pi n/L)$ for the semi-infinite matrix indexed
by the lattice, one has
\begin{equation}
\label{eq.CB}
\sum_{|S|=N,\ S\subset\mathbb{Z}}
\prod_{n\in S}f_n
\det X^{(p)}[S]\det X^{(k)}[S]
=\text{(multilinear expansion of)}\
\det\left[\sum_n f_nX^{(p)}_{n,i}X^{(k)}_{n,j}\right] ,
\end{equation}
with $f_n=\ee^{2\pi i ny/L}$. Keeping track of the fixed border
column $q_0$ (which must appear in both determinants exactly once) and
taking $q_0\to\infty$ produces the difference structure
\begin{equation}
\chi(y)=\det(G+B)-\det G ,
\end{equation}
where $B$ is the rank-one matrix \eref{eq.Bdef} descending from the
border, and
\begin{equation}
\label{eq.Gentry}
G_{ij}= -\frac{2}{L}\sum_{n\in\mathbb{Z}}
\frac{\ee^{2\pi i ny/L}}
{\left(p_i-\frac{2\pi n}{L}\right)\left(k_j-\frac{2\pi n}{L}\right)}
\ee^{-i y(p_i+k_j)/2}
\left|\frac{\partial p_i}{\partial\Lamt}\right|^{1/2}
\left|\frac{\partial k_j}{\partial\Lam}\right|^{1/2} ,
\end{equation}
where the symmetric phase split $\ee^{-i y(p_i+k_j)/2}$ uses the
equality $Q_p=Q_k$ of the total momenta of the two states. For $p_i\neq k_j$, partial fractions turn the lattice sum
into the difference of two sums of the type \eref{eq.efun}, which on
the Bethe roots take the on-shell form and yield \eref{eq.Adef}. For a
shared root $p_i=k_j=\kappa$ the lattice sum in \eref{eq.Gentry} is
instead the $\kappa$-derivative of the sum in \eref{eq.efun},
\begin{equation}
-\frac2L\sum_n\frac{\ee^{2\pi i ny/L}}{(\kappa-2\pi n/L)^{2}}
=\frac2L\partial_\kappa\left[\frac{L}{2}\ee^{i\kappa y}
\left(\cot\frac{\kappa L}{2}-i\right)\right] .
\end{equation}
Because $\partial_\kappa$ acts on $\cot(\kappa L/2)$, the entry
differs from the naive $p_i\to k_j$ limit of \eref{eq.Adef} by
\begin{equation}
-\left[\frac{L}{2}\left(1+(\alpha\kappa-\Lam)^2\right)+\alpha\right]
\frac{\partial k_j}{\partial\Lam}=-1 ,
\end{equation}
by \eref{eq.dkdL}. This proves \eref{eq.shared-entry}. With these
entries \eref{eq.rho-det} is an identity in finite volume.

\section{Small-\texorpdfstring{$y$}{y} expansion and the matrix
elements}
\label{app:smally}

Multiply row $i$ and column $j$ of $A+B$ and of $A$ by, respectively,
\begin{equation}
\ee^{-i yp_i/2}\left|\frac{\partial p_i}{\partial\Lamt}\right|^{-1/2},
\qquad
\ee^{i yk_j/2}\left|\frac{\partial k_j}{\partial\Lam}\right|^{-1/2} .
\end{equation}
The total factor is unity for the phases (equal total momenta) and the
$\partial$-factors are compensated by the prefactor written in
\eref{eq.rho-prefactor} below. This brings $\chi(y)$ to the form
\begin{equation}
\label{eq.rho-prefactor}
\chi(y)=
\frac{\prod_{i}\left|\frac{\partial p_i}{\partial\Lamt}\right|^{1/2}
\prod_{j}\left|\frac{\partial k_j}{\partial\Lam}\right|^{1/2}}
{\sqrt{\sum_i\frac{\partial p_i}{\partial\Lamt}}
\sqrt{\sum_j\frac{\partial k_j}{\partial\Lam}}}
\left[\det\left(\mathcal A+\mathcal B\right)-\det\mathcal A\right] ,
\end{equation}
\begin{equation}
\label{eq.A19}
\begin{gathered}
\mathcal A_{ij}=\frac{\Lamt-\Lam}{p_i-k_j}-\alpha
+\frac{\ee^{i(k_j-p_i)y}-1}{p_i-k_j}(\alpha k_j-\Lam-i),\\
\mathcal B_{ij}=\ee^{-i yp_i} ,
\end{gathered}
\end{equation}
where the normalization of $B$ has been moved into the prefactor of
\eref{eq.rho-prefactor}. Thus $\mathcal B=b\otimes\mathbf 1^{\rm T}$, with $b_i=\ee^{-i yp_i}$ and
$\mathbf 1$ the vector with all entries equal to one, is of rank one at
every $y$, with a constant right vector.

Expanding \eref{eq.A19},
\begin{equation}
\label{eq.A-expansion}
\mathcal A_{ij}=C_{ij}-v_i-w_j+\frac{(i y)^2}{2}\alpha p_ik_j
+O(y^3),
\end{equation}
with $C$ the matrix defined in \srf{sec.matrix-elements},
\begin{equation}
v_i=\alpha-i y(\Lam+i)+\frac{(i y)^2}{2}(\Lam+i)p_i ,
\qquad
w_j=i\alpha yk_j+\frac{(i y)^2}{2}k_j(\alpha k_j-\Lam-i) .
\end{equation}
The term $v_i$, being constant along rows, spans the same rank-one
direction as $\mathcal B$: for any matrix $X$ the function
\begin{equation}
u\mapsto\det(X+u\otimes\mathbf 1^{\rm T})
\end{equation}
is affine in $u$, so the difference of determinants
\begin{equation}
\det(X-v\otimes\mathbf 1^{\rm T}+b\otimes\mathbf 1^{\rm T})
-\det(X-v\otimes\mathbf 1^{\rm T})
\end{equation}
depends on $b$ only, and $v_i$ may be dropped. One is left with
\begin{equation}
\mathcal A=C+i yA_1+(i y)^2A_2,\qquad
\mathcal B\propto b_0+i yb_1+(i y)^2 b_2 ,
\end{equation}
\begin{equation}
[A_1]_{ij}=-\alpha k_j,\quad
[A_2]_{ij}=\frac{\alpha p_ik_j}{2}-\frac{k_j}{2}(\alpha k_j-\Lam-i),
\quad
[b_0]_{ij}=1,\ \ [b_1]_{ij}=-p_i,\ \ [b_2]_{ij}=\frac{p_i^2}{2} .
\end{equation}
$A_1$ and $\mathcal B$ are rank one, $A_2$ is rank two. Using
\begin{equation}
\det(C+X)=\det C\det(1+C^{-1}X)
\end{equation}
and the shorthand
\begin{equation}
\Tr C^{-1}u\otimes s^{\rm T}\equiv s^{\rm T}C^{-1}u ,
\end{equation}
we write
\begin{equation}
\det(\mathcal A+\mathcal B)-\det\mathcal A
=\det C\left[R_0+i yR_1+(i y)^2R_2+o(y^2)\right]
\end{equation}
and find
\begin{align}
R_0&=\Tr C^{-1}b_0 ,\\
R_1&=\Tr C^{-1}b_0\Tr C^{-1}A_1-\Tr C^{-1}b_0C^{-1}A_1+\Tr C^{-1}b_1 ,\\
R_2&=\Tr C^{-1}b_2+\Tr C^{-1}b_0\Tr C^{-1}A_2-\Tr C^{-1}b_0C^{-1}A_2
\nonumber\\
&\quad+\Tr C^{-1}b_1\Tr C^{-1}A_1-\Tr C^{-1}b_1C^{-1}A_1 .
\end{align}
The rank-one structure of $A_1$ and $b_n$ implies
\begin{equation}
\Tr C^{-1}b_0C^{-1}A_1=\Tr C^{-1}b_0\Tr C^{-1}A_1
\end{equation}
and, with
$[\widetilde A_2]_{ij}=\alpha p_ik_j/2$,
\begin{equation}
\Tr C^{-1}b_1C^{-1}A_1=2\Tr C^{-1}b_0\Tr C^{-1}\widetilde A_2 ,
\qquad
\Tr C^{-1}b_1\Tr C^{-1}A_1=2\Tr C^{-1}b_0C^{-1}\widetilde A_2 ,
\end{equation}
\begin{equation}
\Tr C^{-1}b_0C^{-1}\widetilde A_2
=\frac{\alpha}{2}
\left(\sum_{ij}k_i[C^{-1}]_{ij}\right)\left(\sum_{ij}[C^{-1}]_{ij}p_j\right)
=-\frac12\Tr C^{-1}b_1\Tr C^{-1}A_1 ,
\end{equation}
whence
\begin{equation}
\label{eq.R-final}
R_1=\Tr C^{-1}b_1 ,
\qquad
R_2=\Tr C^{-1}b_2-\Tr C^{-1}b_0\Tr C^{-1}\widetilde A_2
+\frac12\Tr C^{-1}b_1\Tr C^{-1}A_1 .
\end{equation}

Since the entries of $C$ depend on the roots only through
$1/(p_i-k_j)$, its inverse admits the contour representation
\begin{equation}
\label{eq.Cinv}
[C^{-1}]_{ij}=\frac{1}{\Lamt-\Lam}
\oint_{k_i}d x\oint_{p_j}d y\,
\frac{1}{x-y}\prod_\alpha\frac{x-p_\alpha}{x-k_\alpha}
\prod_\alpha\frac{y-k_\alpha}{y-p_\alpha} ,
\qquad
\oint\frac{d x}{x}=1 ,
\end{equation}
(each contour encircles only the indicated point), which is verified by
computing $\sum_j[C^{-1}]_{ij}C_{js}$: deforming the $y$-contour to
enclose all $k$'s, the integral vanishes for $i\ne s$ and equals $1$
for $i=s$. Sums against powers follow by deforming both contours to
infinity:
\begin{equation}
\label{eq.Cinv-sums}
\sum_{ij}[C^{-1}]_{ij}p_j^{n}
=\oint_{\mathcal C_p}\frac{d y}{\Lamt-\Lam}y^{n}
\prod_\alpha\frac{y-k_\alpha}{y-p_\alpha}
=\begin{cases}
\dfrac{\sum_jk_j-\sum_jp_j}{\Lamt-\Lam}=0, & n=0,\\[3mm]
\dfrac{1}{n+1}
\dfrac{\sum_jp_j^{n+1}-\sum_jk_j^{n+1}}{\Lamt-\Lam}, & n=1,2 ,
\end{cases}
\end{equation}
where the contour $\mathcal C_p$ encircles all roots $p_j$ and the
$n=0$ case uses equality of total momenta. The analogous sums with
$k_i^m$ on the left follow by symmetry. Assembling
\eref{eq.R-final}--\eref{eq.Cinv-sums} and restoring the prefactor
\eref{eq.rho-prefactor} --- in which $\det C$ combines with the
$\partial$-factors into $D(k,p)$ of \eref{eq.Ddet} --- gives
\begin{equation}
\chi(y)=\frac{D(k,p)}
{\sqrt{\sum_i\frac{\partial p_i}{\partial\Lamt}}
\sqrt{\sum_i\frac{\partial k_i}{\partial\Lam}}}
\left[-i y\frac{E_p-E_k}{\Lamt-\Lam}
+(i y)^2\left(\frac{\sum_ip_i^3{-}\sum_ik_i^3}{6(\Lamt-\Lam)}
+\frac{\alpha}{2}\left(\frac{E_p-E_k}{\Lamt-\Lam}\right)^{2}\right)
+o(y^2)\right] ,
\end{equation}
from which Eqs.~\eqref{eq.P-offdiag}--\eqref{eq.P2-offdiag} follow by
\eref{eq.nq-def}.

For $p=k$ one has $D(k,k)=1$ and, as the $\Lamt\to\Lam$ limit of
\eref{eq.Cinv},
\begin{equation}
[C^{-1}]_{ij}=\delta_{ij}\frac{\partial k_j}{\partial\Lam} ,
\end{equation}
so that
\begin{equation}
\chi(y)=1-i y\frac{S_1}{S_0}
+\frac{(i y)^2}{2}\frac{S_2-\alpha(S_0S_2-S_1^2)}{S_0}+O(y^3) ,
\end{equation}
reproducing \eref{eq.P-diag}.

Taking $\Lam\to-\infty$ at fixed $\Lamt$, the rapidity factors
attached to the roots $k_j$ behave as
\begin{equation}
\frac{\partial k_j}{\partial\Lam}\to\frac{2}{L\Lam^2},\qquad
(\Lamt-\Lam)\left|\frac{\partial k_j}{\partial\Lam}\right|^{1/2}\to\sqrt{\frac2L},\qquad
\left(\sum_j\frac{\partial k_j}{\partial\Lam}\right)^{-1/2}\sim|\Lam|\sqrt{\frac{L}{2(N{+}1)}} ,
\end{equation}
so that $D(k,p)\to D_\infty(k,p)$ of \eref{eq.Dinf}, and the
normalization combines with $(E_p-E_k)/(\Lamt-\Lam)$ into the finite
results \eref{eq.special-offdiag}. The diagonal elements
\eref{eq.special-diag} follow directly from the decomposition
\eref{eq.special-decomp}.

\section{The degenerate sectors at infinite coupling}
\label{app:degenerate}

Let $\alpha=0$ and let $|p\rangle\ne|k\rangle$ have equal rapidity:
common phase shift $\delta$, roots on integer branches
$\{m_i\}\neq\{n_j\}$ with $\sum_im_i=\sum_jn_j$, sharing
$r=N{+}1-u$ roots ($u\ge2$). All derivatives $\partial k/\partial\Lam$
and $\partial p/\partial\Lamt$ take the common value
$d=\frac2L\sin^2\delta$. Order rows and columns so that shared
roots occupy the first $r$ (aligned) slots. Using
\eref{eq.Adef}--\eref{eq.shared-entry} at $\alpha=0$,
\begin{equation}
A_{ij}=-(\Lam+i)d
\frac{2i\sin\left(y(p_i-k_j)/2\right)}{p_i-k_j}
-\left[\Delta_r\right]_{ij},
\qquad
B_{ij}=\frac{\ee^{-i yp_i/2}\ee^{-i yk_j/2}}{N+1} ,
\end{equation}
with $\Delta_r$ the projector on the shared slots (the $\sin$-form
follows from \eref{eq.Adef} because both states have the same
$\Lam$. Its diagonal limit $i y$ agrees with the first term of
\eref{eq.shared-entry}). Note that $B$ is rank one and that
the expansion of $A+\Delta_r$ contains only odd powers of $y$,
with the $y^1$ term proportional to the all-ones matrix $J$ and the
$y^3$ term of rank three.

Expand \eref{eq.rho-det} by multilinearity in columns. Since
$B$ is rank one it contributes exactly once, and each of the $u$
non-shared columns of $A$ starts at order $y$ with one and the same
column vector $\propto\mathbf 1$. At order $y^0$ every term needs all
non-shared columns at their $O(y)$ terms at least, so the coefficient
vanishes ($\chi(0)=0$: orthogonality). At order $y^1$, once the
$B$-column is used, at least $u-1\ge1$ non-shared columns must take
their $O(y)$ term $\propto\mathbf 1$, and the $O(y^0)$ term of the
$B$-column is also $\propto\mathbf 1$. Every term contains two
proportional columns, which gives the first relation in
\eref{eq.selection}. At order
$y^2$ with $u=2$, the only nonvanishing assignments put one non-shared
column at $i yc\mathbf 1$ ($c=-(\Lam+i)d$) and the $B$-column at its
$O(y)$ vector $\propto(p_i)$. The two choices of which non-shared
column takes $\mathbf 1$ enter the determinant with opposite
orientation and cancel. For $u\ge3$ every term is $O(y^3)$ or
smaller, which gives the second relation and completes the proof of
\eref{eq.selection}. For $u=2$ the vanishing of the second relation
holds in the symmetric (Abel) sense discussed in
\srf{sec.matrix-elements}.

Matrix elements of higher powers $\Pimp^{n}$, $n\ge3$, do not
vanish in general in the degenerate sectors (the $y^3$ term of
$A+\Delta_r$ has rank three), which is consistent with the fact that
they are also UV-sensitive.

\section{Determinant representation of \texorpdfstring{$P(t)$}{P(t)}}
\label{app:pt}

Insert \eref{eq.sum-to-int} for both state sums in \eref{eq.Pt-def}
(with variables $\Lam_k,x$ and $\Lam_p,y$), use the overlap
determinants \eref{eq.overlap} for $c_a$ and the matrix elements
\eref{eq.P-offdiag}. In the primitive $\Phi(t)$ of
\srf{sec.dynamics} the summand
factorizes: the two determinants from the overlaps and the one
from $D(k,p)$ are minors of matrices indexed by the same lattice momenta,
so the subset summation of \arf{app:cauchy-binet} applies,
and the sums over both branch configurations $\{n_j\}$ collapse into a
single bordered $(N{+}1)\times(N{+}1)$ determinant \cite{GPZ2015},
\begin{equation}
\label{eq.Bt-detA}
\Phi(t)=\iint d\Lam_k\,d\Lam_p
\iint\frac{d x\,d y}{(2\pi)^2}
\ee^{i(y-x)p_0}
\det\begin{pmatrix}
\mathsf A_{ij}&\ell\mathsf v_i\\[1mm]
\mathsf w_j&\Gamma
\end{pmatrix},
\qquad \ell=\frac{2(\Lam_k-\Lam_p)}{L} ,
\end{equation}
where $i,j$ run over the $N$ occupied momenta $q_i$ of the Fermi sea
and the entries are double sums over all Bethe branches
$k=k(\Lam_k)$, $p=p(\Lam_p)$:
\begin{align}
\label{eq.entries-discrete}
\mathsf A_{ij}&=\ell\sum_{k,p}
\frac{\ee^{i t(p^2-k^2)/2+i x(k-q_j)-i y(p-q_i)}}
{(p-q_i)(p-k)(k-q_j)}
\frac{\partial k}{\partial\Lam_k}\frac{\partial p}{\partial\Lam_p},
&
\mathsf v_i&=\sum_{k,p}
\frac{\ee^{i t(p^2-k^2)/2+i xk-i y(p-q_i)}}
{(p-q_i)(p-k)}
\frac{\partial k}{\partial\Lam_k}\frac{\partial p}{\partial\Lam_p},
\nonumber\\
\mathsf w_j&=\sum_{k,p}
\frac{\ee^{i t(p^2-k^2)/2+i x(k-q_j)-i yp}}
{(p-k)(k-q_j)}
\frac{\partial k}{\partial\Lam_k}\frac{\partial p}{\partial\Lam_p},
&
\Gamma&=\sum_{k,p}
\frac{\ee^{i t(p^2-k^2)/2+i xk-i yp}}{p-k}
\frac{\partial k}{\partial\Lam_k}\frac{\partial p}{\partial\Lam_p} .
\end{align}
The bordered determinant obeys the identity
\begin{equation}
\label{eq.border-identity}
\det\begin{pmatrix}\mathsf A&\ell \mathsf v\\
\mathsf w&\Gamma\end{pmatrix}
=(\Gamma-1)\det\mathsf A
+\det(\mathsf A-\ell\mathsf v\otimes \mathsf w) ,
\end{equation}
which follows from the matrix determinant lemma,
\begin{equation}
\det(\mathsf A-\ell\mathsf v\otimes\mathsf w)
=\det\mathsf A-\ell\mathsf w^{\rm T}{\rm adj}(\mathsf A)\mathsf v .
\end{equation}

The virtue of \eref{eq.entries-discrete} is that time derivatives
factorize: $-2i\partial_t$ brings down $(p-k)(p+k)$ from the phase,
\begin{equation}
-2i\partial_t\ee^{i t(p^2-k^2)/2}=(p-k)(p+k)\ee^{i t(p^2-k^2)/2} ,
\end{equation}
the factor $p-k$ cancels the denominator, and $p+k$ splits the double
sums,
\begin{equation}
\label{eq.dtA}
-2i\partial_t\mathsf A_{ij}
=\ell\left[\widetilde E_1(q_i)E_0(q_j)+\widetilde E_0(q_i)E_1(q_j)\right],
\qquad
-2i\partial_t\Gamma=E_0\widetilde E_1+\widetilde E_0E_1 ,
\end{equation}
and similarly for $\mathsf v,\mathsf w$, with the one-index sums
\begin{equation}
\label{eq.En}
E_n(q)=\sum_k\frac{k^n\ee^{-i tk^2/2+i x(k-q)}}{k-q}
\frac{\partial k}{\partial\Lam_k},
\quad
E_n=\sum_kk^n\ee^{-i tk^2/2+i xk}\frac{\partial k}{\partial\Lam_k},
\end{equation}
and $\widetilde E_n(q),\widetilde E_n$ their $(t,x,\Lam_k)\to
(-t,-y,\Lam_p)$ counterparts. Likewise, the $x$- and $y$-derivatives
of $\mathsf A$ are rank-one matrices built from $\mathsf v$ and
$\mathsf w$,
\begin{equation}
-i\partial_x\mathsf A_{ij}=\ell\mathsf v_i\ee^{-i q_jx},
\qquad
i\partial_y\mathsf A_{ij}=\ell\ee^{i yq_i}\mathsf w_j .
\end{equation}

Because every derivative is a sum of rank-one updates, derivatives of
the determinants reduce to finite differences. For any matrix
$\mathsf A$ and vectors $u_a,s_a$ define
\begin{equation}
C_{\mathsf A}(u_1,u_2;s_1,s_2)\equiv
\det(\mathsf A+\ell u_1\otimes s_1+\ell u_2\otimes s_2)
-\det(\mathsf A+\ell u_1\otimes s_1)
-\det(\mathsf A+\ell u_2\otimes s_2)+\det \mathsf A ,
\end{equation}
which is linear in each argument and antisymmetric under
$u_1\leftrightarrow u_2$ at fixed $s$'s (and vice versa). With
$e_n(q_j)=E_n(q_j)$, $\tilde e_n(q_i)=\widetilde E_n(q_i)$ regarded
as vectors on the Fermi sea, the time derivative of
\eref{eq.border-identity} assembles into
\begin{equation}
\begin{aligned}
&(-2i\partial_t)\left[(\Gamma-1)\det\mathsf A
+\det(\mathsf A-\ell \mathsf v\otimes \mathsf w)\right]\\
&\quad=-C_{\mathsf A}(\mathsf v,\tilde h;\mathsf w,e_0)
-C_{\mathsf A}(\mathsf v,\tilde e_0;\mathsf w,h)
+i(\partial_y-\partial_x)\left[E_0\widetilde E_0\det\mathsf A\right]\\
&\qquad+i\partial_y\left[\det(\mathsf A-\ell \mathsf v\otimes e_0\widetilde
E_0)-\det\mathsf A\right]
-i\partial_x\left[\det(\mathsf A-\ell\tilde e_0\otimes \mathsf w
E_0)-\det\mathsf A\right]\\
&\qquad+\det\left(\mathsf A+\ell\tilde h\otimes(\Gamma e_0-\mathsf w E_0)\right)
+\det\left(\mathsf A+\ell(\Gamma\tilde e_0-\mathsf v\widetilde E_0)
\otimes h\right)-2\det\mathsf A ,
\end{aligned}
\end{equation}
where the vectors
\begin{equation}
h(q)=e_1(q)+\ee^{-i xq}E_0,
\qquad
\tilde h(q)=\tilde e_1(q)+\ee^{i yq}\widetilde E_0
\end{equation}
collect the boundary terms of the $x$- and $y$-integrations by parts,
and
$\dot\Gamma$ is given by \eref{eq.dtA}. The $x$- and $y$-derivative
terms are then traded for factors of $p_0$ by integration by parts,
\begin{equation}
\int d x\,\ee^{-i xp_0}\partial_x[\cdots]
=i p_0\int d x\,\ee^{-i xp_0}[\cdots] .
\end{equation}
Finally, the involution $I$, complex conjugation combined with
$(x,y,\Lam_k,\Lam_p)\to(y,x,\Lam_p,\Lam_k)$, acts as
\begin{equation}
\mathsf A_{ij}\to-\mathsf A_{ji},\quad \mathsf v\to-\mathsf w,\quad
e_n\to\tilde e_n,\quad E_n\to\widetilde E_n,\quad \ell\to-\ell,\quad
\Gamma\to-\Gamma,\quad h\to\tilde h,
\end{equation}
leaves $\ee^{i p_0(x-y)}$ invariant, and maps the integrand to its
complex conjugate, halves the number of independent structures,
and makes $P(t)$ manifestly real, giving \eref{eq.Pt-final}. The
intermediate step is
\begin{equation}
\label{eq.Bt-final}
\begin{gathered}
\Phi(t)=\iint d\Lam_kd\Lam_p\iint
\frac{d xd y}{(2\pi)^2}\ee^{i(y-x)p_0}
\left[(\Gamma-1)\det(1+\hat K)+\det(1+\hat K-\delta\hat K_0)\right] ,\\
\delta K_0(q,q')=\frac{\Lam_k-\Lam_p}{\pi}v(q)w(q') .
\end{gathered}
\end{equation}

Sums over Bethe branches are evaluated by the residue formula
\begin{equation}
\label{eq.residue}
\sum_k\frac{\partial k}{\partial\Lam_k}f(k)
=\frac{1}{2\pi i}\oint_{\mathcal C}
\frac{d k\,f(k)}{\alpha k-\Lam_k-\cot(kL/2)} ,
\end{equation}
valid because $\partial k/\partial\Lam$ is the reciprocal of
$\partial_k[\alpha k-\Lam-\cot(kL/2)]$ on shell. Observe that the contour $\mathcal C$ encircles
all roots but no singularities of $f$. Deforming $\gamma$ to the real
axis shifted by $\pm i0$ and using
$\cot(kL/2)\to\mp i$ for $\operatorname{Im}k\to\pm\infty$ (at
$L\to\infty$), each branch sum becomes an explicit principal-value
integral plus, for functions with poles at Bethe roots, a residue
term. In this way the entries \eref{eq.entries-discrete} become, per
row and column, functions of the free momenta $q\in[-\kF,\kF]$
only, and $\det$ becomes a Fredholm determinant with measure
$\frac{L}{2\pi}\int_{-\kF}^{\kF}d q$, where the factor $\ell$ combines
with the measure into $(\Lam_k-\Lam_p)/\pi$. The scalar functions in
\eref{eq.kernels} are
\begin{align}
\label{eq.v-int}
v(q)&=\int_{-\infty}^{\infty}\frac{d z}{2\pi i}
\left[\frac{\ee^{i tz^2/2-i y(z-q)}}{(\alpha z-\Lam_p-i)(z-q-i0)}
-\frac{\ee^{i tz^2/2-i y(z-q)}}{(\alpha z-\Lam_p+i)(z-q+i0)}
\right]\sigma(z) ,\\
\label{eq.sigma-int}
\sigma(z)&=-\int_{-\infty}^{\infty}\frac{d z'}{2\pi i}
\left[\frac{\ee^{-i tz'^2/2+i xz'}}
{(\alpha z'-\Lam_k-i)(z'-z-i0)}
-\frac{\ee^{-i tz'^2/2+i xz'}}
{(\alpha z'-\Lam_k+i)(z'-z+i0)}\right]
+\frac{\ee^{-i tz^2/2+i xz}}{\Lam_p-\Lam_k} ,\\
\label{eq.EEt-int}
E(q)&=\int\frac{d z}{2\pi i}
\left[\frac{\ee^{-i tz^2/2+i x(z-q)}}
{(\alpha z-\Lam_k-i)(z-q-i0)}
-\frac{\ee^{-i tz^2/2+i x(z-q)}}
{(\alpha z-\Lam_k+i)(z-q+i0)}\right] ,
\end{align}
the functions $\Et(q)$ and $\widetilde\sigma$ follow from $E(q)$ and
$\sigma$ by
\begin{equation}
(t,x,\Lam_k)\to(-t,-y,\Lam_p)
\qquad\text{and}\qquad
(t,x,\Lam_k)\leftrightarrow(-t,-y,\Lam_p),
\end{equation}
respectively, $w(q)$ is built like $v(q)$ with $\sigma$ replaced by
$\widetilde\sigma$, and
\begin{equation}
\label{eq.E0-int}
\begin{gathered}
E_0=\int_{-\infty}^{\infty}\frac{d k}{\pi}
\frac{\ee^{-i tk^2/2+i kx}}{(\alpha k-\Lam_k)^2+1},
\qquad
\Et_0=\int_{-\infty}^{\infty}\frac{d p}{\pi}
\frac{\ee^{i tp^2/2-i py}}{(\alpha p-\Lam_p)^2+1},\\
\Gamma=\int\frac{d z}{\pi}
\frac{\ee^{i tz^2/2-i yz}\sigma(z)}{(\alpha z-\Lam_p)^2+1} .
\end{gathered}
\end{equation}
The double $z$-integrals in $v,w,\Gamma$ can be reduced to single
integrals by inserting the explicit pole decomposition of $\sigma$.

\section{Infinite-coupling kernels and their limits}
\label{app:alpha0}

At $\alpha=0$ all denominators in
Eqs.~\eqref{eq.v-int}--\eqref{eq.E0-int} lose their $z$-dependence
through $\alpha z$, the $z$-integrals become Gaussian-times-rational,
and one obtains Eqs.~\eqref{eq.v-erf}--\eqref{eq.gauss-factors} using
\begin{equation}
\label{eq.vp-erf}
\mathrm{v.p.}\int\frac{d k}{\pi}
\frac{\ee^{-i tk^2/2+i kx}}{k-q}
=i\ee^{-i tq^2/2+i qx}
\Erf\left(\frac{(x-qt)(1-i)}{2\sqrt t}\right) ,
\end{equation}
and its conjugate.

As $t\to0^+$ the error functions with arguments built on $x$ and
$y$ saturate to $\sgn(x)$ and $\sgn(y)$, while
\begin{equation}
V\to-i\sgn(x{-}y)\sgn(x),\qquad W\to-i\sgn(x{-}y)\sgn(y)
\end{equation}
(the remaining integrals are $O(\sqrt t)$ by stationary phase). One finds
\begin{equation}
\label{eq.t0-kernels}
\begin{gathered}
v(q)=-\frac{\ee^{i xq}}{\Lam_p+i\sgn(x{-}y)}
\frac{\Lam_p+i\sgn(x)}{(\Lam_k+i\sgn(x))(\Lam_p-\Lam_k)},\\
w(q)=-\frac{\ee^{-i yq}}{\Lam_k+i\sgn(x{-}y)}
\frac{\Lam_k-i\sgn(y)}{(\Lam_p-i\sgn(y))(\Lam_p-\Lam_k)},\\[1mm]
E(q)=-\frac{1}{\Lam_k+i\sgn(x)},\qquad
\Et(q)=-\frac{1}{\Lam_p-i\sgn(y)},\qquad
E_0=\frac{2\delta(x)}{1+\Lam_k^2},\qquad
\Et_0=\frac{2\delta(y)}{1+\Lam_p^2},
\end{gathered}
\end{equation}
and the kernel assembles into
\begin{equation}
\label{eq.K-t0}
\begin{aligned}
\left.K(q,q')\right|_{t=0}
={}&\frac{\ee^{i x(q-q')}-1}{\pi(q-q')}
\frac{\Lam_p+i\sgn(x)}{(\Lam_k+i\sgn(x))(\Lam_p+i\sgn(x{-}y))}\\
&-\frac{\ee^{i y(q-q')}-1}{\pi(q-q')}
\frac{\Lam_k-i\sgn(y)}{(\Lam_p-i\sgn(y))(\Lam_k+i\sgn(x{-}y))} ,
\end{aligned}
\end{equation}
a two-parameter deformation of the counting (Fisher--Hartwig) kernel.
The emergence of
$P(0)=p_0$ from the $\delta$-collapsed term is shown in
\eref{eq.P0-emergence}.

For $q,x,y=O(1)$ the error functions tend to $-\sgn(q)$ and,
rescaling $k\to k/t$ in $V(q)$,
\begin{equation}
V(q)\to\frac2\pi\int_{|x-y|}^{\infty}\frac{d u}{u}
\ee^{-i u\sgn(x-y)q}+i\sgn(q) ,
\end{equation}
so that (for $x>y$; the opposite ordering follows by the involution)
\begin{equation}
\label{eq.tinf-limits}
v(q)\to\frac{\ee^{i xq}}{(1{+}\Lam_k^2)(\Lam_p+i)}
\left[\frac{\Lam_k\Lam_p+1}{\Lam_k-\Lam_p}-i\sgn(q)
+\frac{2i}{\pi(\Lam_p-i)}\int_{x-y}^{\infty}\frac{d u}{u}
\ee^{-i uq}\right]+O(t^{-1/2}) ,
\end{equation}
with the analogous limit for $w(q)$, and
\begin{equation}
\begin{gathered}
E(q)\to-\ee^{-i tq^2/2}\;\frac{\Lam_k+i\sgn(q)}{\Lam_k^2+1},
\qquad
\Et(q)\to-\ee^{i tq^2/2} \; \frac{\Lam_p-i\sgn(q)}{\Lam_p^2+1},\\
E_0,\Et_0=O(t^{-1/2})\to0 .
\end{gathered}
\end{equation}

\end{appendix}

\bibliography{refs}

\end{document}